\documentclass[11pt]{article}

\usepackage{epsf}
\usepackage{amsmath}
\usepackage{amssymb}
\usepackage{graphicx,epsfig}
\usepackage{dcolumn}
\usepackage{bm}

\begin{document}

\def\thefootnote{\fnsymbol{footnote}}

\vspace*{-1.5cm}
\begin{flushright}
\tt MTA-PHYS-0704
\end{flushright}

\vspace{0.2cm}
\begin{center}
\Large\bf\boldmath
Estimating the annihilation decay $B_s\to\rho\gamma$ with factorization
\unboldmath
\end{center}

\vspace{0.4cm}
\begin{center}
M. R. Ahmady$^{\,1,}$\footnote{Electronic address: {\tt mahmady@mta.ca}} and F. Mahmoudi$^{\,2,3,}$\footnote{Electronic address: \tt mahmoudi@in2p3.fr}\\[0.4cm]
{\sl $^1$ Department of Physics, Mount Allison University, 67 York Street, Sackville,\\
 New Brunswick, Canada E4L 1E6\\
\vspace{0.3cm}
$^2$ High Energy Physics, Uppsala University, Box 535, 75121 Uppsala, Sweden\\
\vspace{0.3cm}
$^3$ Clermont Universit\'e, Universit\'e Blaise Pascal, CNRS/IN2P3,\\ LPC, BP 10448, 63000 Clermont-Ferrand, France\\}
\end{center}
\vspace{0.5cm}

\begin{abstract}
The branching ratio for the rare two-body $B_s\to\rho\gamma$ decay is calculated using the factorization assumption. This transition is dominated by the annihilation diagrams and, in principle, prone to receiving substantial contributions from new physics.  We estimate $Br(B_s\to\rho\gamma ) \approx 1.6 \times 10^{-9}$ within the Standard Model and investigate the sensitivity of this decay mode to the effects of two new physics scenarios: vector quark model and supersymmetry. Our results indicate that the shift in branching ratio is at most around $10\%$ with the addition of vector quarks and is negligibly small in the constrained minimal supersymmetric extension of the Standard Model.
\end{abstract}
\vspace{0.3cm}

\section{Introduction}
With Large Hadron Collider (LHC) becoming operational, search for new physics beyond the Standard Model (SM) is entering an exciting new phase. For example, among other particles, a large number of $B_s$ mesons are expected to be produced at this facility, and as a result, rare processes involving this interesting hadronic system will be accessible experimentally. The transition $B_s\to\rho\gamma$ is of particular interest as, unlike $B\to\rho\gamma$, it mainly proceeds via the so-called annihilation process. As such, it is interesting to study the stability of the SM prediction for the branching ratio of this decay mode to the effects of, as yet unknown, exotic particles.\\
\\
In this paper, we first calculate the branching ratio for $B_s\to\rho\gamma$ decay within the SM using the light cone amplitudes and factorization assumption, after identifying the effective operators which have the dominant role in this process. To the best of our knowledge, even though our prediction has a large uncertainty, this is the first calculation of the branching ratio for this transition. Next, we check the sensitivity of this decay mode to possible tree-level flavor changing neutral currents (FCNC) generated by new physics. One particular scenario is the addition of a single down-type vector quark to the three quark generations of the SM. Since both the left-handed and the right-handed components of a vector quark are iso-singlets, the weak interactions of this quark can only happen via mixing with the ordinary quarks. Consequently, the extended $3\times 4$ quark mixing matrix is not unitary and tree-level FCNC is present in this model. In fact, one can show that the non-zero $b\to sZ^\circ$ in Vector Quark Model (VQM) is proportional to $U^{sb}=V_{us}^*V_{ub}+ V_{cs}^*V_{cb}+V_{ts}^*V_{tb}$ which basically measures the non-closure of the unitarity triangle within the SM. Our results indicate that, within the acceptable range for the above model parameter, the shift from the SM prediction is about $10\%$.\\
\\
We also consider the minimal supersymmetric extension of the Standard Model where every fundamental particle receives a supersymmetric partner with spin differing by 1/2 unit. The particle spectrum of supersymmetry therefore consists of three neutral Higgs bosons, two charged Higgs bosons and supersymmetric particles such as gluino, squarks, sleptons, neutralinos and charginos. This leads to additional contributions to the $B_s\to\rho\gamma$ branching ratio due to the new particles. We study in this paper a scenario with minimal flavor violation and evaluate the numerical implications of such contributions. We show that the overall supersymmetric contributions in this decay mode are very small.\\
\\
We would like to emphasize again that unlike some well known decay channels like the inclusive $b\to s\gamma$, exclusive $B\to K^*\gamma$ and $B_s\to \Phi\gamma$ which have been under thorough investigation in the literature, $B_s\to\rho\gamma$ proceeds mainly via annihilation process and in that sense is quite different and unique.
\section{Low energy effective Hamiltonian}
\noindent  The relevant effective Hamiltonian for $B_s\to\rho\gamma$ can be written as
\begin{equation}
{\cal H}_{eff}=\frac{G_F}{\sqrt{2}}\sum_{p=u,c} V^*_{ps}V_{pb} [ C_1(\mu) \, O^p_1(\mu) + C_2(\mu) \, O^p_2(\mu)
+ \sum_{i=3}^8 C_i(\mu) \,
O_i(\mu)]\;, \label{heffective}
\end{equation}
where $C$'s are the Wilson coefficients and the operators are defined as follows:
\begin{eqnarray}
\nonumber O_1^p&=&\bar s_\alpha\gamma^\mu P_Lp_\beta\bar p_\beta\gamma_\mu P_Lb_\alpha \; ,\raisebox{0cm}[0cm][0.4cm]{~}\\
\nonumber O_2^p&=&\bar s_\alpha\gamma^\mu P_L p_\alpha\bar p_\beta\gamma_\mu P_Lb_\beta \; ,\raisebox{0cm}[0cm][0.3cm]{~}\\
\nonumber O_3&=&\bar s_\alpha\gamma^\mu P_Lb_\alpha\sum_{q}\bar q_\beta\gamma_\mu P_Lq_\beta \; ,\\
\nonumber O_4&=&\bar s_\alpha\gamma^\mu P_Lb_\beta\sum_{q}\bar q_\beta\gamma_\mu P_Lq_\alpha \; ,\\
\nonumber O_5&=&\bar s_\alpha\gamma^\mu P_Lb_\alpha\sum_{q}\bar q_\beta\gamma_\mu P_Rq_\beta \; ,\\
O_6&=&\bar s_\alpha\gamma^\mu P_Lb_\beta\sum_{q}\bar q_\beta\gamma_\mu P_Rq_\alpha \; , 
\end{eqnarray}%
\begin{eqnarray}
\nonumber O_7 &=&\frac{e}{4\pi^2} m_b \bar s_\alpha\sigma^{\mu\nu} P_R b_\alpha F_{\mu\nu} \raisebox{0cm}[0cm][0.4cm]{~}\; ,\\
\nonumber O_8 &=&\frac{g_s}{4\pi^2} m_b \bar s_\alpha\sigma^{\mu\nu} P_R T^a_{\alpha\beta}b_\beta G_{\mu\nu}^a \raisebox{0cm}[0cm][0.2cm]{~}\;.
\label{operators}
\end{eqnarray}%
\begin{figure}[t!]
\centerline{$\begin{array}{cc}
\epsfig{file=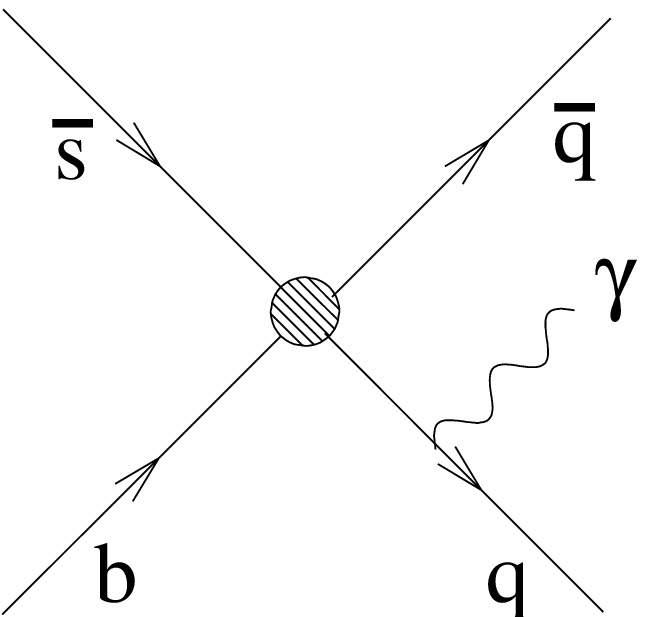,width=2.5cm}\qquad&\epsfig{file=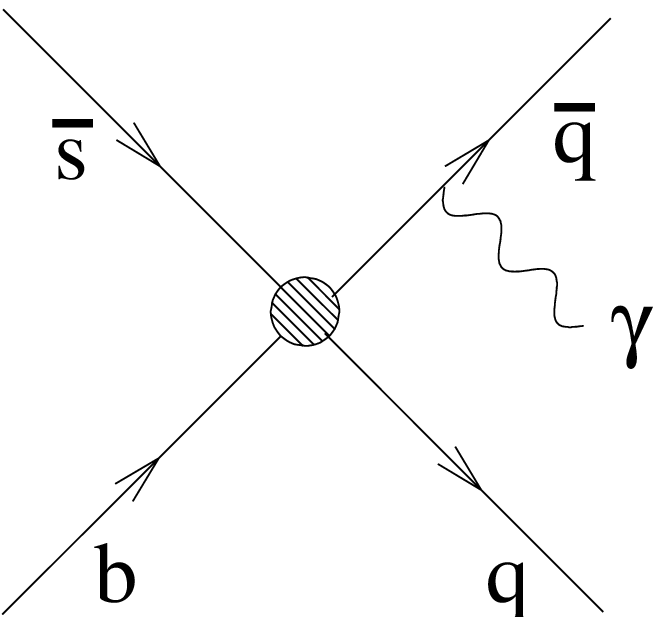,width=2.5cm}\\[0.5cm]
\epsfig{file=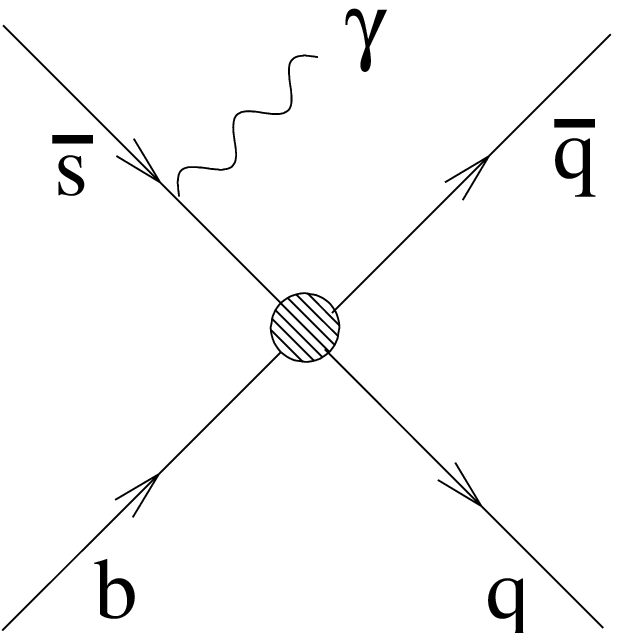,width=2.5cm}\qquad&\epsfig{file=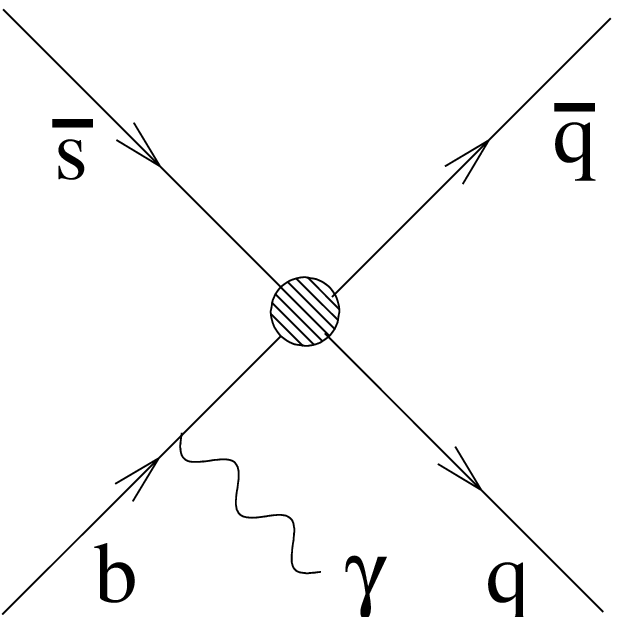,width=2.5cm}\end{array}$}
\vspace*{8pt}
\caption{Annihilation diagrams contributing to the $B_s\to\rho\gamma$ decay. The first row corresponds to $\mathcal{A}_1$ (left) and $\mathcal{A}_2$ (right) and the second row corresponds to $\mathcal{A}_3$ (left) and $\mathcal{A}_4$ (right).\protect\label{diag}}
\end{figure}
The contributions of the current-current operators $O_1^u$ and $O_2^u$ and QCD penguin operators $O_i$, $i=3\ldots 6$ are shown in Figure \ref{diag}. Operators $O_1^c$, $O_2^c$, $O_7$ and $O_8$, on the other hand, have vanishing amplitudes to the leading order within the factorization assumption. As a result, the amplitude for the $B_s\to\rho\gamma$ decay is composed of four generic matrix elements of the following forms:
\begin{eqnarray}
\mathcal{A}_1 &=& \langle \rho(k_\rho, \epsilon_\rho) \gamma(k_\gamma, \epsilon_\gamma) | \bar{s} \gamma^\mu (1-\gamma_5) b \, \bar{q} \gamma_\mu (1\pm\gamma_5)
\frac{1}{\not{\! p}_1-m_q} \gamma_\nu \epsilon_\gamma^\nu q| B_s(p_{B_s}) \rangle \;,
\label{A1}\\
\mathcal{A}_2 &=& \langle \rho(k_\rho, \epsilon_\rho) \gamma(k_\gamma, \epsilon_\gamma) | \bar{s} \gamma^\mu (1-\gamma_5) b \, \bar{q}  \gamma_\nu
\epsilon_\gamma^\nu \frac{1}{\not{\! p}_2-m_q} \gamma_\mu (1\pm\gamma_5) q| B_s(p_{B_s}) \rangle \;,
\label{A2}
\\
\mathcal{A}_3 &=& \langle \rho(k_\rho, \epsilon_\rho) \gamma(k_\gamma, \epsilon_\gamma) | \bar{s} \gamma_\nu \epsilon_\gamma^\nu \frac{1}{\not{\! p}_3-m_s} \gamma^\mu (1-\gamma_5) b \, \bar{q}  \gamma_\mu
 (1\pm\gamma_5) q| B_s(p_{B_s}) \rangle \;,
\label{A3}
\\
\mathcal{A}_4 &=& \langle \rho(k_\rho, \epsilon_\rho) \gamma(k_\gamma, \epsilon_\gamma) | \bar{s} \gamma^\mu (1-\gamma_5)  \frac{1}{\not{\! p}_4-m_b} \gamma_\nu \epsilon_\gamma^\nu b \, \bar{q}  \gamma_\mu
 (1\pm\gamma_5) q| B_s(p_{B_s}) \rangle \;,
\label{A4}
\end{eqnarray}
where $p_1=k_\gamma + p_q$, $p_2=k_\gamma + p_{\bar{q}}$, $p_3=p_{\bar{s}} - k_\gamma$, and $p_4= p_b - k_\gamma$, with $q=u\; {\rm and}\;d$. \\
\\
The contributions from the amplitudes ${\cal A}_{1,2,4}$ are power suppressed as compared to ${\cal A}_3$ \cite{bosch,ali01}. However, ${\cal A}_{1,2}$ give a numerically important power correction because of a relative enhancement factor from the Wilson coefficients.\\
The leading twist light-cone amplitude for a vector meson ($V$) with flavor content ($\bar{q} q'$) and polarization vector $\epsilon_V$ can be expressed as \cite{ballbraun98,ballbraun96}:
\begin{eqnarray}
\nonumber\langle V(k_V, \epsilon_V) | \bar{q}(0)_\alpha q'(z)_\beta |0\rangle \Bigr|_{z^2=0} &=& - \frac{f^\perp_V}{8} [\not{\! \epsilon}_V,\not{\! k}_V]_{\beta\alpha} \int_0^1 dv e^{i\bar{v} k_V \cdot z} \Phi_V^\perp(v) \nonumber \\
&&  - \frac{f_V m_V}{4}\left[ i \not{\! k}_{\beta\alpha} (\epsilon_V \cdot z) \int_0^1 dv e^{i\bar{v} k_V \cdot z} \Phi_V^\parallel(v) \right. \label{matrixelement}\\
&& \nonumber + (\not{\! \epsilon}_V)_{\beta\alpha} \int_0^1 dv e^{i\bar{v} k_V \cdot z} g_V^{\perp\,(v)}(v)\\
&& \nonumber  \left.- \frac{1}{4} \bigl(\epsilon_{\kappa\lambda\mu\nu} \epsilon_V^\kappa k_V^\lambda z^\mu \gamma^\nu \gamma_5\bigr)_{\beta\alpha} \int_0^1 dv
e^{i\bar{v} k_V \cdot z} g_V^{\perp\,(a)}(v) \right] \;,
\end{eqnarray}
where the quark and antiquark contents of the vector meson have momenta $p_{q'} = v k_V$ and $p_{\bar{q}} = \bar{v} k_V$, respectively, with $\bar{v}=1-v$.
The functions $f(v)=\Phi_V^\perp(v)$, $\Phi_V^\parallel(v)$, $g_V^{\perp\,(v)}(v)$ and $g_V^{\perp\,(a)}(v)$ are all normalized, such as $\displaystyle \int_0^1 dv f(v) =1$.\\
\\
The relevant distribution amplitudes for the $\rho$ meson in $B_s\to\rho\gamma$ are given by \cite{ballbraun98}:
\begin{equation}
\phi^\perp_V(v) =  6 v\bar v \left[ 1 + 3 a_1^\perp\, \chi + a_2^\perp\, \frac{3}{2} ( 5\chi^2  - 1 ) \right]\;,
\label{DA1}
\end{equation}
\begin{eqnarray}
g_V^{\perp\,(v)}(v) &=& \frac{3}{4}(1+\chi^2) + a_1^\parallel\,\frac{3}{2}\, \chi^3 + \left(\frac{3}{7} \, a_2^\parallel + 5 \zeta_{3} \right) \left(3\chi^2-1\right) \nonumber\\
&& + \left[ \frac{9}{112}\, a_2^\parallel + \frac{15}{64}\, \zeta_{3}\Big(3\,\omega_{3}^V-\omega_{3}^A\Big)
 \right] \left( 3 - 30 \chi^2 + 35\chi^4\right) \\
&& +\frac{3}{2}\,\widetilde{\delta}_+\,(2+\ln v + \ln\bar v) + \frac{3}{2}\,\widetilde{\delta}_-\, ( 2 \chi + \ln\bar v - \ln v)\;,\nonumber
\label{DA2}
\end{eqnarray}
with $\chi = v - \bar{v} = 2v-1$.  Other coefficients in (\ref{matrixelement}), (\ref{DA1}) and (\ref{DA2}) can be found in Tables \ref{coeff1} and \ref{coeff2}.
Note that in the case of a $\rho$-meson, $\widetilde{\delta}_+$ and $\widetilde{\delta}_-$ vanish. \\
\\
The light-cone distribution amplitude for $B_s$ meson to leading power can be written as \cite{bosch,beneke2000,ali07}:
\begin{eqnarray}
&& \langle 0 | \bar{s}(z)_\alpha \, b(0)_\beta |B_s(p_{B_s}) \rangle \Bigr|_{z_+=z_\perp =0} =
 [(\not{\! p_{B_s}}+m_{B_s}) \gamma_5]_{\beta\gamma} \nonumber \\
&& \times \frac{i f_{B_s}}{4} \int_0^1 d\xi e^{-i\xi p_{B_s+} z_-}
[\Phi_{B_s1}(\xi)+\not{\! n_-}\Phi_{B_s2}(\xi)]_{\gamma\alpha} \label{matrixelement2}
\end{eqnarray}
where $n_-=(1,0,0,-1)$, with the light cone components for any vector $v$ defined as:
\begin{equation}
v_\pm = \frac{v^0 \pm v^3}{\sqrt{2}} \;.
\end{equation}
The light spectator carries longitudinal momentum fraction $\xi = p_{\bar{s}+}/p_{B_s+}$.
The normalization conditions are
\begin{equation}
\int_0^1 d\xi \Phi_{B_s1}(\xi) =1\;, \hspace{0.5cm} \int_0^1 d\xi \Phi_{B_s2}(\xi) =0\;.
\end{equation}
The contribution of $\Phi_{B_s2}$ is numerically small \cite{lu} and is neglected in the following. We parameterize the first negative moment of $\Phi_{B_s1}$ by the quantity $\lambda_{B_s}$, such as
\begin{equation}
\int_0^1 d\xi \frac{\Phi_{B_s1}(\xi)}{\xi} = \frac{m_{B_s}}{\lambda_{B_s}} \;.
\end{equation}
Using the model function from \cite{ali07}, we evaluated the value of $\lambda_{B_s} = 580 \pm 190$ GeV.
The numerical values for the mass parameters which are used in our calculations are given in Table \ref{masses}.\\
\\
\begin{table}
\renewcommand{\arraystretch}{1.4}
\addtolength{\arraycolsep}{3pt}
$$
\begin{array}{|c|c|c|c|c|}\hline
f_\rho \mbox{ (MeV)}& f^\perp_\rho \mbox{ (MeV)}& a_1^\parallel,a_1^\perp & a_2^\parallel &  a_2^\perp \\ \hline 198 \pm 7 & 152 \pm 9 & 0& 0.16 \pm 0.1& 0.17
\pm 0.1 \\ \hline
\end{array}
$$
\caption[]{Decay constants and couplings of distribution amplitudes for $\rho$-meson, including SU(3)-breaking. The renormalization scale is $\mu^2=4.8$
GeV$^2$.} \label{coeff1}
$$
\begin{array}{|c|c|c|c|c|c|c|}\hline
\zeta_3 & \omega_3^A & \omega_3^V & \omega_3^T & \zeta_4 & \zeta_4^T & \tilde{\zeta_4^T}\\ \hline 0.023& -1.8& 3.7& 7.5& 0.13& 0.07& -0.07
\\ \hline
\end{array}
$$
\caption[]{Couplings for distribution amplitudes without including SU(3)-breaking. The renormalization scale is $\mu^2=4.8$ GeV$^2$.} \label{coeff2}
\renewcommand{\arraystretch}{1}
\addtolength{\arraycolsep}{-3pt}
$$
\begin{array}{|c|c|c|}\hline
m_q & m_\rho & m_{B_s} \\ \hline 3 \mbox{ MeV} & 775 \mbox{ MeV} & 5367 \mbox{ MeV}\\ \hline
\end{array}
$$
\caption[]{Quark and meson masses used in our calculations.} \label{masses}
\renewcommand{\arraystretch}{1}
\addtolength{\arraycolsep}{-3pt}
\end{table}%
Consequently, from the formulas (\ref{A1}) to (\ref{A4}), the decay rate can be obtained as:
\begin{equation}
\Gamma(B_s\to \rho \gamma)=\frac{\alpha}{8} \frac{m_{B_s}^2-m_\rho^2}{m_{B_s}^3} G_F^2 |V^*_{ts} V_{tb}|^2 |\mathcal{A}|^2 \;,\label{decayrate}
\end{equation}
where $|\mathcal{A}|^2$ is given as:
\begin{equation}
|\mathcal{A}|^2 \equiv |C_{12}|^2 |\mathcal{A}_{1,2}|^2+|C_{34}|^2 |\mathcal{A}_{3,4}|^2+2 \mbox{Re} ( C_{12} \, C_{34}^*\, \mathcal{A}_{1,2} \, \mathcal{A}_{3,4}^*) \;, \label{Asquare}
\end{equation}
with $\mathcal{A}_{1,2}$ and $\mathcal{A}_{3,4}$ defined as the following:
\begin{equation}
\mathcal{A}_{1,2} \equiv \mathcal{A}_1-\mathcal{A}_2 = \pm \, 2 f_{B_s} f_\rho m_\rho  (I_1-I_2) \epsilon_{\kappa\lambda\mu\nu} \, \epsilon_\rho^\kappa \, p_{B_s}^\lambda \, k_\rho^\mu \, \epsilon_\gamma^\nu \;, \label{A12}
\end{equation}
\begin{eqnarray}
\mathcal{A}_{3,4} \equiv \mathcal{A}_3-\mathcal{A}_4 &=& - \frac {i f_{B_s} f_\rho m_\rho}{2 E_\gamma \lambda_{B_s}} \Bigl\{ \bigl[ (\epsilon_\rho \cdot k_\gamma) \, (p_{B_s} \cdot \epsilon_\gamma) - \, (\epsilon_\rho \cdot \epsilon_\gamma) \, (p_{B_s} \cdot k_\gamma) \bigr] \nonumber \\
&& \times \bigl(1-\frac{\lambda_{B_s}}{m_{B_s}}\bigr) 
+ \, i \, \epsilon_{\kappa\lambda\mu\nu} \, \epsilon_\rho^\kappa \, p_{B_s}^\lambda \, \epsilon_\gamma^\mu \, k_\gamma^\nu \, \bigl(1+\frac{\lambda_{B_s}}{m_{B_s}}\bigr) \Bigr\} \; .
\label{A34}
\end{eqnarray}
In Eq. (\ref{A12}), plus and minus signs correspond to $V+A$ and $V-A$ structure, respectively, and $E_\gamma$ in Eq.~(\ref{A34}) is the energy of the photon. Eqs. (\ref{matrixelement}) and (\ref{matrixelement2}) are used to parameterize the hadronic matrix elements.  Accordingly, $I_1$ and $I_2$ in Eq.~(\ref{A12}) are defined as:
\begin{eqnarray}
I_1&=&\int_0^1 \frac{g^{\perp\,(v)}_\rho(v)}{v m_{B_s}^2+ v \bar{v} m_\rho^2 - m_q^2} \;, \label{I1}\\
I_2&=&\int_0^1 \frac{v \, g^{\perp\,(v)}_\rho(v)}{v m_{B_s}^2+ v \bar{v} m_\rho^2 - m_q^2} \;.
\end{eqnarray}
Neglecting the small light quark mass, one can show that:
\begin{eqnarray}
I_1&=& \frac{1}{m_{B_s}^2}\bigl(-5.22 \pm 0.33 + (2.47 \pm 0.10) X \bigr) \;,\\
I_2&=& \frac{0.99 \pm 0.01}{m_{B_s}^2} \;,
\end{eqnarray}%
where the logarithmically divergent integral $\int_0^1 dv/v$ in Eq.~(\ref{I1}) has been parameterized by $X=(1+\rho e^{i \varphi}) \ln(m_{b}/\Lambda_h)$ with $\Lambda_h \approx 0.5$ GeV a typical hadronic scale, $\phi$ an arbitrary strong-interaction phase, and $\rho \le 1$ \cite{beneke2000}.\\
$C_{12}$ and $C_{34}$ in Eq.~(\ref{Asquare}) are combinations of Wilson coefficients appearing in (\ref{heffective}):
\begin{eqnarray}
C_{12} &\equiv& \frac{1}{\sqrt{2}}\left[\frac{2}{3}  \left( C_1+ \frac{C_2}{3}\right) \frac{V^*_{us} V_{ub}}{V^*_{ts} V_{tb}} + \left( C_3 +  \frac{C_4}{3} - C_5 - \frac{C_6}{3}\right) \right] \;,
\label{c12}\\
C_{34} &\equiv& \frac{1}{\sqrt{2}}\left[\frac{1}{3}  \left( C_1+ \frac{C_2}{3}\right) \frac{V^*_{us} V_{ub}}{V^*_{ts} V_{tb}} \right] \;.
\label{c34}
\end{eqnarray}
The Wilson coefficients are real, however the CKM matrix elements can have imaginary parts which are taken into account. The numerical values for $C_{12}$ and $C_{34}$ within the SM are calculated to be $C_{12}^{SM}=\bigl((0.7 \pm 0.1) + i (8.7 \pm 0.9)\bigr) \times 10^{-3}$ and $C_{34}^{SM}=\bigl((-1.7 \pm 0.1) + i (4.4 \pm 0.2)\bigr) \times 10^{-3}$ using the inputs and conventions of \cite{wilsoncoeff}.  The errors in these coefficients are mainly due to scale dependence and determination of the CKM matrix elements.\\
\\
Our result for the branching ratio can then be expressed in the following form:
\begin{equation}
Br(B_s\to \rho \gamma) = (1.6^{+0.6}_{-0.8}) \times 10^{-9} \, \left(\frac{f_{B_s}}{230 \mbox{ MeV}}\right)^2 \left(\frac{V_{ts}V_{tb}}{0.041}\right)^2 \;.
\label{bratio}
\end{equation}
The large uncertainties in the above result are mainly due to the acute sensitivity of our predictions to the parametrization of the divergent integral, and therefore this result must be considered as a rough estimate for this branching ratio.\\
\\
The expected production yield of a decay mode per year at LHCb can be estimated from the following formula \cite{lhcb}:
\begin{equation}
N_{\rm signal}=2\sigma_{b\bar b}\,\mathcal{L}^{int}_{\rm year}\times Br(\bar b\to B_s)\times Br({\rm decay\;\; mode})\;,\label{yield}
\end{equation}
where $\sigma_{b\bar b}$, the $b\bar b$ production cross section is expected to be about 500 $\mu b$, the integrated luminosity for one year of data taking is
$\mathcal{L}^{int}_{\rm year}=2$ fb$^{-1}$ with the assumption of an average luminosity of $2\times 10^{32}$ cm$^{-2}$s$^{-1}$, and the branching ratio of b
quark(antiquark) hadronizing into a $\bar B_s(B_s)$ meson is taken to be
\begin{equation}
Br(\bar b\to B_s)=Br(b\to\bar B_s)=(10.4\pm 1.4)\times 10^{-2}\; .
\end{equation}
Based on Eq. (\ref{yield}) and our estimated branching ratio in Eq. (\ref{bratio}), one would expect about 340 $B_s(\bar B_s)\to\rho\gamma$ per year.  The signature of this decay mode is two pions with invariant mass around $\rho$ resonance plus a photon.  The key in successful measurement of this decay channel could be the high energy resolution of the calorimeter so to make it possible to make the required background cut which is expected to be mainly due to $B\to\rho\gamma$ process. The smallness of the likelihood of this process as expected by the SM could make it quite sensitive to the new physics, especially those which allow FCNC at the tree level. We examine one such beyond the SM scenarios as well as supersymmetric contributions in the next section.
%
\section{Contributions from Vector Quark Model and Supersymmetry}
\noindent In the vector-quark model (VQM) \cite{VQM}, in which SM is augmented by an additional isosinglet down type quark the $3\times 4$ quark mixing matrix is not unitary, which leads to non-vanishing FCNC at the tree level. For example, the elementary $b\to s Z^\circ$ vertex is proportional to the non-unitary parameter $U^{sb}=(V^\dag V)^{sb}$ at the leading order which directly measures the non-closure of a unitary triangle and as such, it is quite desirable to find observables which are most sensitive to this parameter. In fact, one can show that the factor $C_{12}$, as defined in (\ref{c12}), receives additional contributions \cite{ac}:
\begin{equation}
\delta C_{12} =\frac{1}{\sqrt{2}}\left[\frac{2}{3}(C_3^{VQM\,up} - C_5^{VQM\,up}) + \frac{1}{3}(C_3^{VQM\,down} - C_5^{VQM\,down}) \right] \;,
\end{equation}
where
\begin{eqnarray}
C_3^{VQM}&=&\frac{U^{sb}}{V_{tb}V_{ts}^*}(I_W^q-Q_q\sin^2{\theta}) \\  &=&\frac{U^{sb}}{V_{tb}V_{ts}^*}\displaystyle\left\{^{\displaystyle \frac{1}{2}-\frac{2}{3}\sin^2{\theta}
=0.35\ldots q=up}_{\displaystyle -\frac{1}{2}+\frac{1}{3}\sin^2{\theta}=-0.42\ldots q=down}\right. \nonumber \\
C_5^{VQM}&=&-\frac{U^{sb}}{V_{tb}V_{ts}^*}Q_q\sin^2{\theta} \\
&=&\frac{U^{sb}}{V_{tb}V_{ts}^*}\displaystyle\left\{^{\displaystyle -\frac{2}{3}\sin^2{\theta}=-0.15\ldots
q=up}_{\displaystyle \frac{1}{3}\sin^2{\theta}=0.08\ldots q=down}\right. \; . \nonumber
\label{cvqm}
\end{eqnarray}%
One can use the experimental measurement of the dileptonic $B\to X_s\ell^+\ell^-$ decay by BELLE \cite{belle} and BABAR \cite{babar} to constrain the acceptable domain for the magnitude and phase of $U^{sb}$. It turns out that this parameter can attain a maximum value of $1.1\times 10^{-3}$ if its phase is close to zero \cite{usbupdate}. The resulting shift in the $C_{12}$ coefficient in this model is at most $\delta C_{12}^{VQM}=3.1\times 10^{-3}$, leading to a branching fraction of about $1.8 \times 10^{-9}$. Therefore the SM branching ratio in (\ref{bratio}) is increased by around $10\%$ with the addition of vector quarks.\\
\\
We also consider supersymmetric contributions to the $B_s\to \rho \gamma$ branching ratio.
In particular diagrams involving charged Higgs boson, charginos and squarks can modify the Wilson coefficients and consequently the branching ratio. To study the numerical implications of such contributions we restricted ourselves to the constrained minimal supersymmetric Standard Model (CMSSM) -- often called mSUGRA -- with the number of free parameters reduced to five assuming the unification at GUT scale. In this scenario, the chargino-squark contributions appearing in the loops are completely negligible. The charged Higgs boson on the other hand can replace the $W$ boson but its contribution is still smaller than the SM contributions. In fact, within the acceptable range for the 5 parameters of mSUGRA, the upper bound for $\delta C^{\rm mSUGRA}$ is smaller than the SM contribution by 3 orders of magnitude and therefore, one can safely assume that the prediction in (\ref{bratio}) is not affected by the constrained SUSY particles.
\section{Conclusion}
\noindent
The rare decay $B_s\to \rho \gamma$ could be an interesting process to look for at the LHC. It may be an experimental challenge to measure this decay mode due to the fact that the final state particles are both neutral. However, knowing the branching ratio helps to devise a strategy for eventual experimental measurement.
\subsection*{Acknowledgments}
\noindent This research is partly supported by a Discovery Grant from NSERC.


\begin{thebibliography}{}
\bibitem{usbupdate} This is the latest update on $U^{sb}$ that will appear in a separate article.
\end{thebibliography}
\end{document}